\documentclass[
reprint,
superscriptaddress,
 amsmath,amssymb,
 aps,
 prl,
 longbibliography
]{revtex4-2}
\usepackage[bookmarks,colorlinks,citecolor=blue,linkcolor=blue]{hyperref}
\usepackage{graphicx}% Include figure files
\usepackage{dcolumn}% Align table columns on decimal point
\usepackage{bm,amssymb,bbold}% bold math
\newcommand{\gdot}{\dot{\gamma}}
\newcommand{\sigmas}{\sigma_{xy}}
\newcommand{\sigman}{\sigma_{yy}}
\newcommand{\Sigmas}{\Sigma_{xy}}
\newcommand{\Sigman}{\Sigma_{yy}}
\newcommand{\phim}{\phi_{\rm m}}
\newcommand{\tauI}{\tau_{\rm I}}
\newcommand{\tauV}{\tau_{\rm V}}

\begin{document}

\title{Nonmonotonic constitutive curves and shear banding in dry and wet granular flows}
\author{Christopher Ness}
\affiliation{School of Engineering, University of Edinburgh, Edinburgh EH9 3FG, United Kingdom}
\author{Suzanne M. Fielding}
\affiliation{Department of Physics, Durham University, Science Laboratories, South Road, Durham DH1 3LE, United Kingdom}

\date{\today}

\begin{abstract}
We use particle simulations to map comprehensively the shear rheology of dry and wet granular matter comprising particles of finite stiffness, in both fixed pressure and fixed volume protocols. At fixed pressure we find non-monotonic constitutive curves that are shear thinning, whereas at fixed volume we find non-monotonic constitutive curves that are shear thickening. We show that the presence of one non-monotonicity does not imply the other. Instead, there exists a signature in the volume fraction measured under fixed pressure that, when present, ensures 
non-monotonic constitutive curves at fixed volume.
In the context of dry granular flow we show that gradient and vorticity bands arise under fixed pressure and volume respectively,
as implied by the constitutive curves. For wet systems our results are consistent with a recent experimental observation of shear thinning at fixed pressure. We furthermore predict discontinuous shear thickening in the absence of critical load friction.
\end{abstract}

\maketitle

Dense granular packings, both dry and suspended in liquid, are among the most abundant materials on earth. They are relevant to manifold  geophysical phenomena, e.g., landslides and debris flows~\cite{iverson1997physics,hutter2005geophysical}, 
and to industrial processes such as
paste extrusion~\cite{o2017critical,prabha2021recent}. 
Understanding their deformation and flow properties is thus of major practical importance. It is also of fundamental interest in  statistical physics, fluid mechanics and rheology~\cite{guazzelli2018rheology,forterre2008flows,jaeger1996granular,ness2022physics}. 

For any complex fluid, a key rheological fingerprint is the constitutive relation  of shear stress $\sigmas$ as a function of shear rate $\gdot$ in  stationary homogeneous shear. For granular materials, there exist two paradigmatic protocols for characterising this relation.
In the first (``fixed volume")~\cite{olsson2007critical,otsuki2009critical,chialvo2012bridging,de2015local,nordstrom2010microfluidic,guy2015towards},
one fixes the sample volume $V$ and measures $\sigmas$ (and sometimes the normal stress or normal stress differences) as a function of $\gdot$ at a given particle volume fraction $\phi$.
In the second (``fixed pressure"),
one fixes the external pressure (usually in fact the normal particle stress ${\sigman}$; but see~\cite{srivastava2021viscometric}) and allows $\phi$ to vary while measuring $\sigmas$ as a function of $\gdot$. 

In the fixed-$V$ protocol, 
dry systems display distinct quasistatic, intermediate and inertial flow regimes dependent on $\gdot$ and $\phi$~\cite{chialvo2012bridging}.
In suspensions the inertial regime is replaced by a viscous regime~\cite{trulsson2012transition,ness2015flow}.
Hysteresis is often observed close to the jamming volume fraction $\phim$ that  
marks the transition between quasistatic and inertial or viscous regimes,
suggesting that the constitutive curve $\sigmas(\gdot)$ has a shear-thickening, S-shaped non-monotonicity.
In the unstable region ${\sigmas}$ can adopt multiple values at a single ${\gdot}$~\cite{grob2014jamming,saw2020unsteady,otsuki2011critical},
implying a predisposition to vorticity banding~\cite{olmsted2008perspectives}.

In the fixed-$\sigman$ protocol, experimental data for the macroscopic friction  $\mu={\sigmas}/{\sigman}$ across a range of scaled shear rates (``inertial number'') $I={\gdot}a/\sqrt{{\sigman}/\rho}$~\cite{jop2006constitutive} 
suggests universal constitutive relations $\mu(I)$ and $\phi(I)$  in dry systems~\cite{gdr2004dense,fall2015dry}.
Here $a$ and $\rho$ are the particle radius and density.
Recent innovations enabling fixed-${\sigman}$ measurements in suspensions~\cite{boyer2011unifying,etcheverry2023capillary}
remarkably suggest analogous relations $\mu(J)$ and $\phi(J)$,
with $J = \eta{\gdot}/{\sigman}$ now the ``viscous number'', where $\eta$ is the solvent viscosity~\cite{boyer2011unifying,tapia2019influence}.
In slow shear (small $I, J$),
$\mu$ was shown to decrease with increasing $I$ in dry simulations~\cite{degiuli2017friction} and experiments~\cite{dijksman2011jamming} (see also~\cite{jaeger1990friction,mills2008rheology,mowlavi2021interplay,da2002viscosity}), and likewise suggested to decrease with increasing $J$ in experiments on wet systems~\cite{perrin2019interparticle}, before increasing at large $I, J$. This gives a shear-thinning  non-monotonicity, in which $\gdot$ can adopt multiple values at a single $\mu$,
implying a predisposition to gradient banding.

%==========================
%==========================
\begin{figure*}
\includegraphics[trim = 0mm 43mm 0mm 0mm, clip,width=0.96\textwidth,page=1]{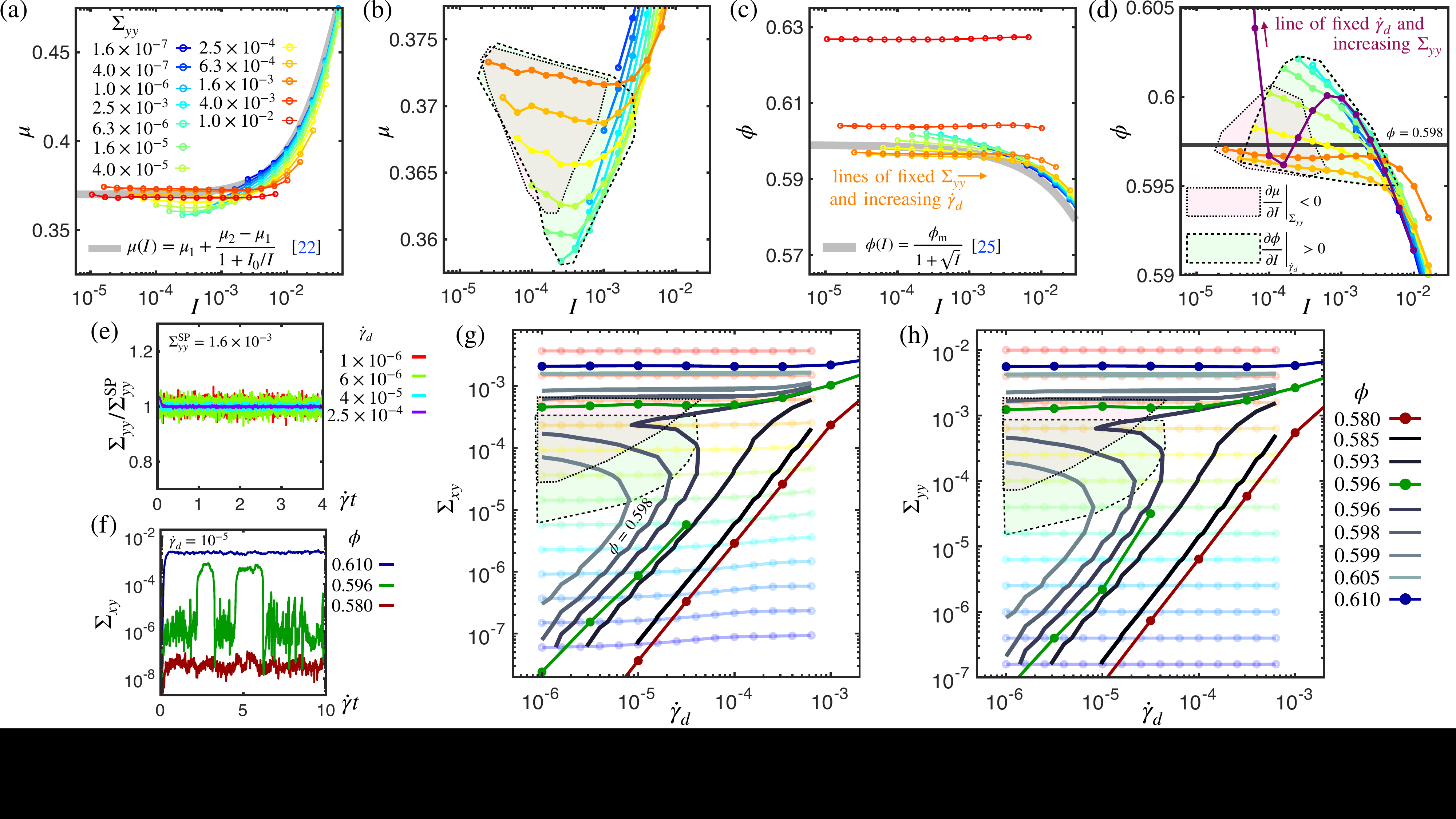}
\caption{
Non-monotonic constitutive curves of dry granular material
at fixed $\Sigman$ (a)-(e)
and fixed $V$ (f)-(h).
Shown are (a) $\mu(I)$ and (c) $\phi(I)$, respectively zoomed in (b), (d).
Each curve of color blue to red has increasing $\gdot_d$ left to right at a fixed $\Sigman$ shown by the legend in (a).
Purple line in (d) connects points with ${\gdot}_d=4\times10^{-6}$ and
varying ${\Sigman}$.
Broad grey lines show fits of~\cite{jop2006constitutive} (a)
and~\cite{boyer2011unifying} (c).
Our fixed-$\Sigman$ feedback algorithm maintains a set-point $\Sigman^\mathrm{SP}=1.6\times10^{-3}$ over time (e).
In (f) are fixed-$V$ time series of $\Sigmas$ at $\gdot_d=10^{-5}$
for various $\phi$.
In (g),(h) we first replot fixed-$\Sigman$ data from (a)-(d) as faded lines
blue to red, then 
connect points of equal
$\phi$ on these curves by lines grey to black.
This gives 
fixed-$V$ constitutive curves, reconstructed from the fixed-$\Sigman$ data.
Shown by darker red, green and blue lines in (g),(h) are constitutive curves measured actually at 
fixed $V$, by binning time series of
$\Sigmas$ (shown in (f)) and $\Sigman$ as described in the text.
Shaded boxes in (b),(d),(g),(h) cover regimes
that have
$\frac{\partial \mu}{\partial I}\vert_{\Sigma_{yy}}<0$ (pink)
and 
$\frac{\partial \phi}{\partial I}\vert_{\gdot_d}>0$ (green),
transcribed across fixed-$\Sigman$ and fixed-$V$ protocols.
}
\label{figure1}
\end{figure*}
%==========================
%==========================

In this Letter, we advance the understanding of granular rheology in three key directions.
First, we demonstrate, within a single model granular system, constitutive curves that are non-monotonic and shear thinning (thickening) at fixed $\sigman$ (fixed $V$),
and show a mapping whereby data can be transposed between these two representations. Significantly, we find that non-monotonicity at fixed $\sigman$ does not imply the same at fixed $V$, 
uncovering instead a signature in the fixed-$\sigman$ curves $\phi(I)$ (or $\phi(J)$)  that, if present, implies non-monotonic $\sigmas(\gdot)$ at fixed $V$.
Second, we show in the context of dry granular systems shear bands with layer-normals in the gradient direction at fixed $\sigman$; and vorticity direction at fixed $V$, revealing a predisposition toward heterogeneous flow consistent with the measured constitutive curves.
Third, we provide the first simulation evidence for non-monotonic $\mu(J)$ in suspensions, recently suggested experimentally~\cite{perrin2019interparticle}, and show that discontinuous shear thickening (DST) can arise at fixed $V$ even in the absence of critical load friction~\cite{seto2013discontinuous,wyart2014discontinuous}, hitherto considered prerequisite.

%==========================
%==========================
\emph{Simulation.---}
%==========================
%==========================
We simulate the Newtonian dynamics of a packing of spheres of density $\rho$
with bidisperse radii $a$ and $1.4a$ (chosen to prevent crystallization~\cite{o2003jamming}) using LAMMPS~\cite{ness2023simulating,thompson2022lammps,cheal2018rheology}.
Interparticle contacts are modelled as Hookean with stiffness $k$, frictional
with sliding coefficient $\mu_p=0.5$, and damped with normal and tangential restitution coefficient 0.5.
We verified that varying $k$ over an order of magnitude does not significantly change any of our results.
For wet systems we also implement pairwise lubrication forces.
Between any two particles $\alpha$ and $\beta$
the leading term for particle $\alpha$ is 
$F^\alpha_i = \frac{\kappa}{h}n_in_j(u^\beta_j-u^\alpha_j)$ with $\kappa$ a scalar function of the radii~\cite{jeffrey1992calculation},
$n$ the centre-to-centre unit vector, $u^\alpha$ and $u^\beta$ the particle velocities and $h$ the surface-to-surface distance. (Roman  suffices denote Cartesian directions.)
The force is truncated when $h<10^{-3}a$.
For details see~\cite{cheal2018rheology}.

\begin{figure}
\includegraphics[trim = 0mm 0mm 0mm 0mm, clip,width=0.48\textwidth,page=1]{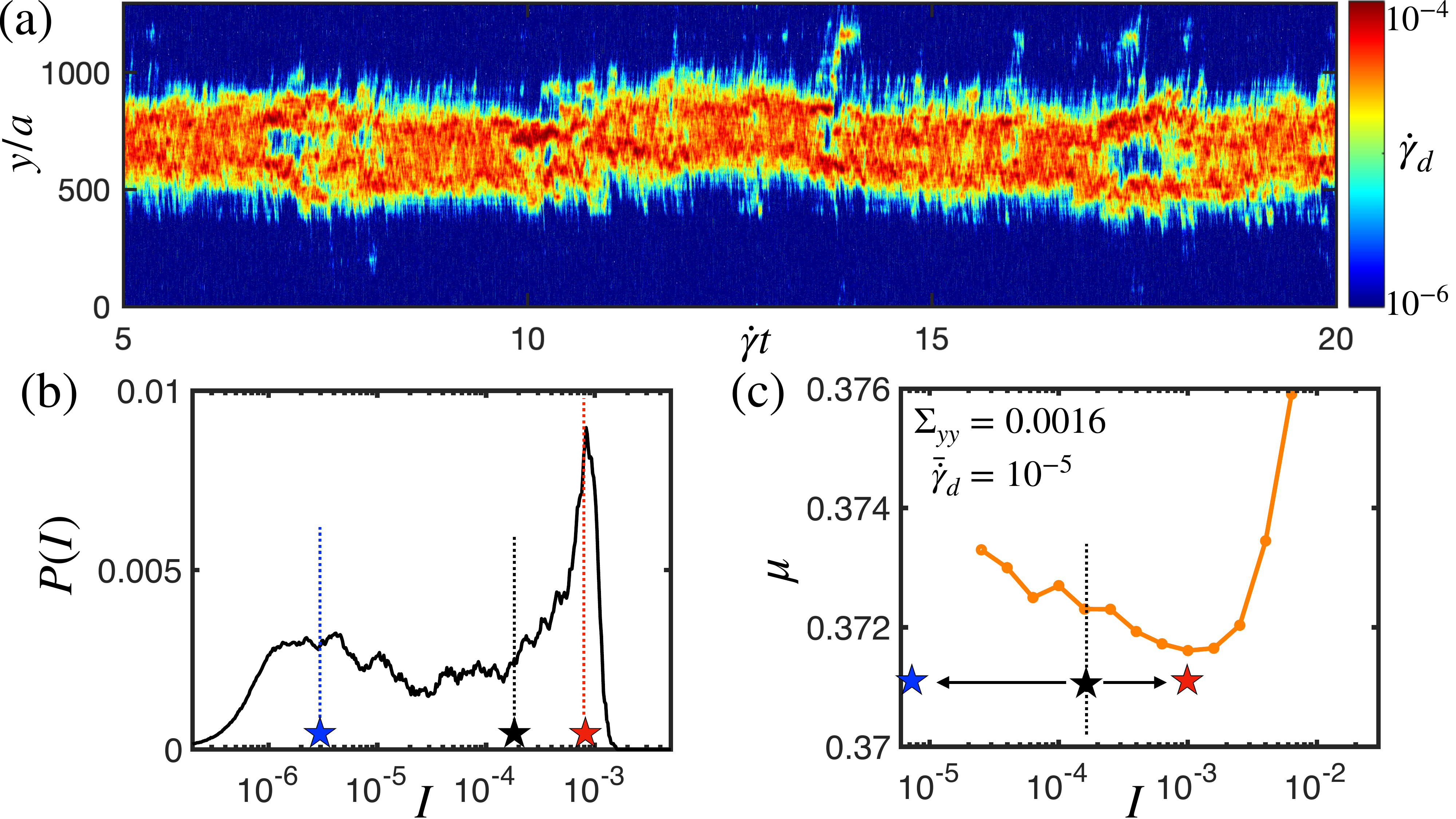}
\caption{
Gradient banding in dry system.
(a) Strain series of $\gdot_d$ profile across $y$,
measured at fixed global ${\Sigman}=0.0016$ and $\bar{\gdot}_d=10^{-5}$. 
Binning temporal and spatial $I$ data gives the
histogram in (b). From its
peaks we obtain $I$ in the low (blue star) and high (red star) shear regions.
These are then plotted as $\mu(I)$ in (c).
Orange line in (c) represents the homogeneous $\mu(I)$ at $\Sigman=0.0016$ from Fig.~\ref{figure1}(a).
Dashed black lines in (b),(c) indicate imposed global $I$.
}
\label{figure2}
\end{figure}

\begin{figure}
\includegraphics[trim = 0mm 0mm 0mm 0mm, clip,width=0.48\textwidth,page=1]{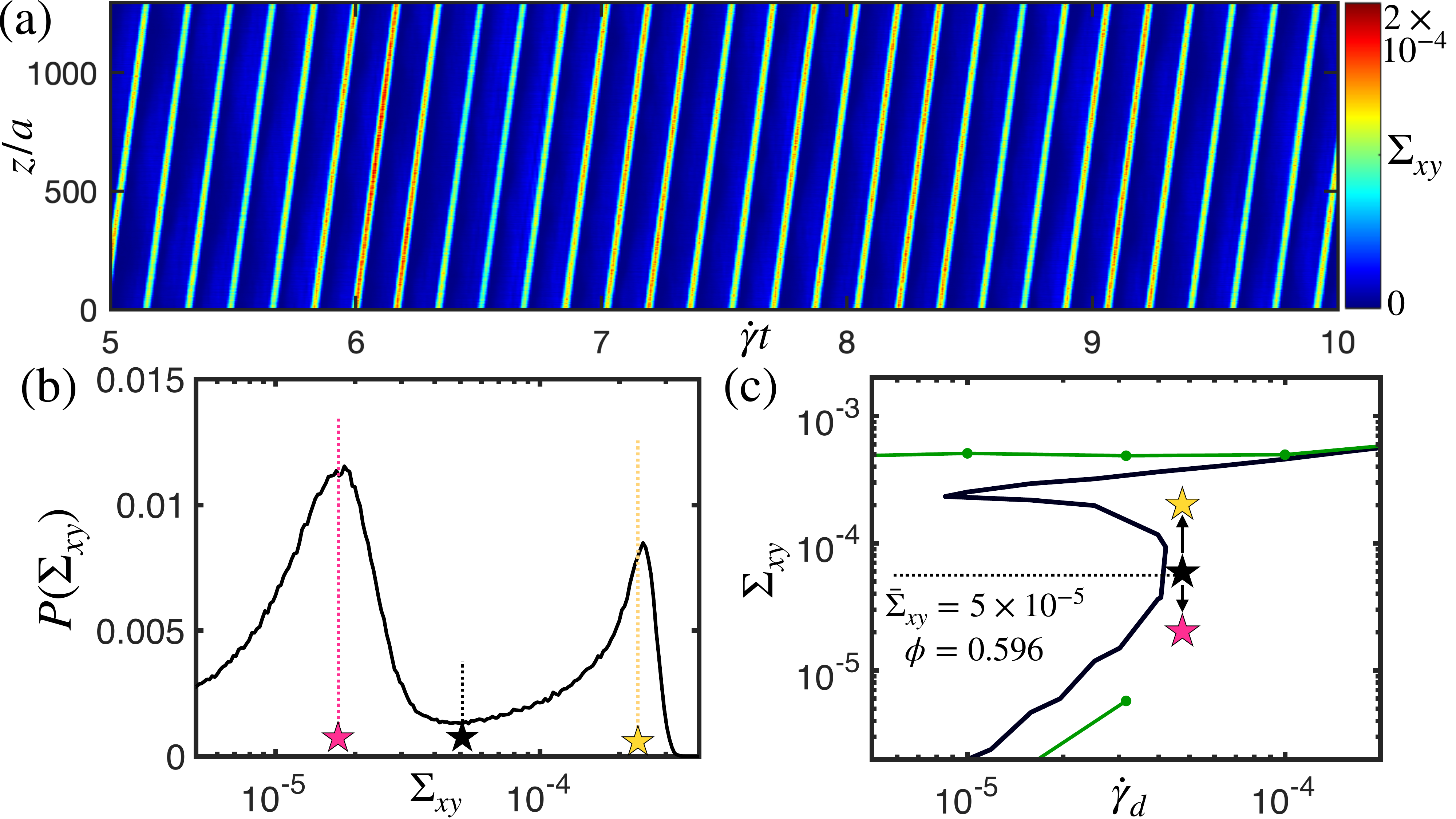}
\caption{
Vorticity banding in dry system.
(a) Strain series of $\Sigmas$ profile across $z$ measured at fixed global $\phi=0.596$ and $\bar{\Sigma}_{xy}=5\times10^{-5}$.
Binning temporal and spatial $\Sigmas$ leads to the
histogram in (b).
From its peaks we obtain the low (pink star)
and high (yellow star) $\Sigmas$ values,
plotted in (c) with those obtained under fixed $V$ and under the reconstruction (green and black
lines, Fig.~\ref{figure1}(g)) at the same $\phi$.
Dashed black lines in (b), (c) indicate imposed global $\bar{\Sigma}_{xy}$.
}
\label{figure3}
\end{figure}

\begin{figure}
\includegraphics[trim = 0mm 0mm 242mm 0mm, clip,width=0.47\textwidth,page=1]{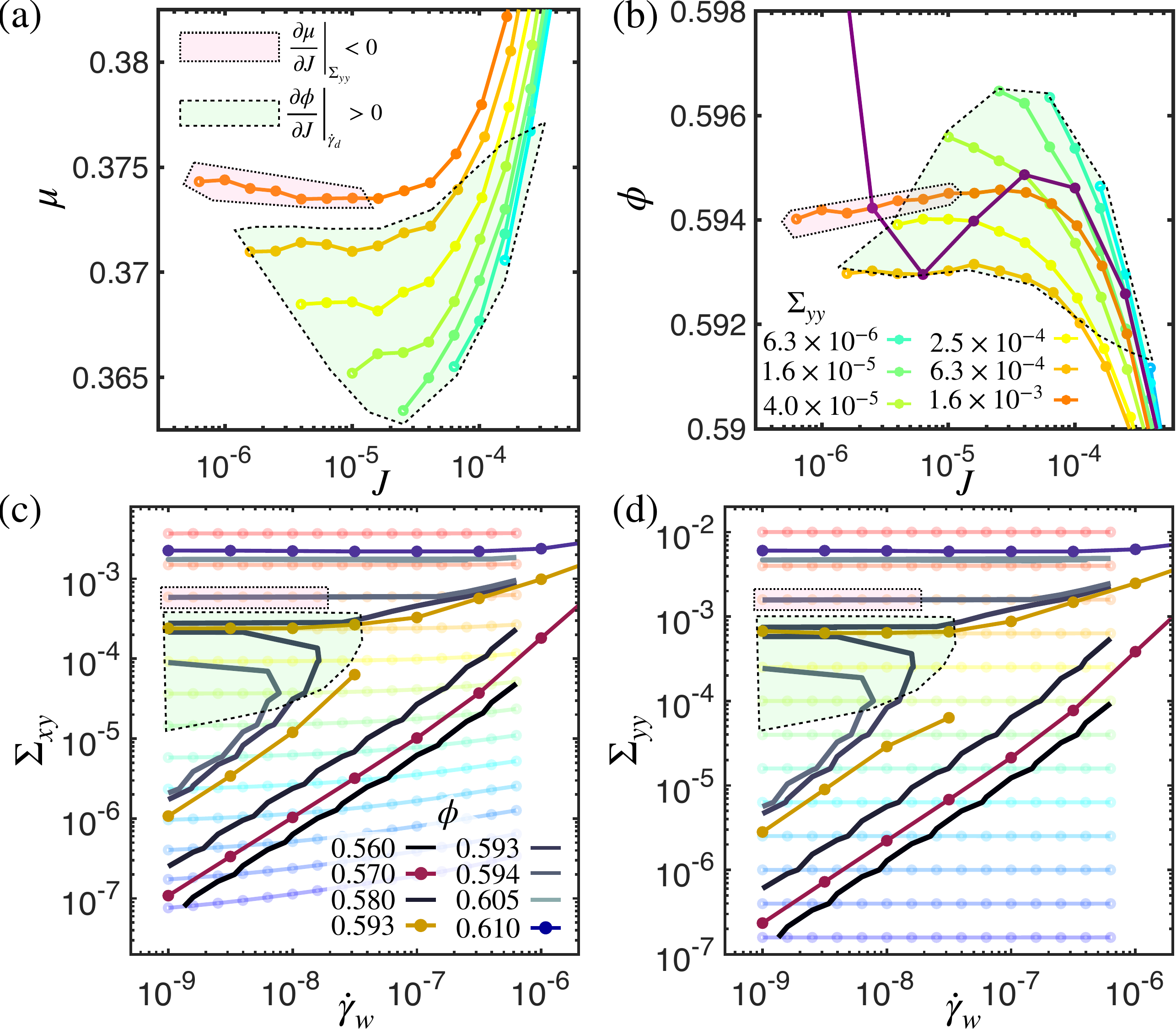}
\caption{
Non-monotonic constitutive curves of wet granular material.
Shown in (a) and (b) are $\mu(J)$ and $\phi(J)$;
in (c) and (d) are the same data
along with reconstructed and actual fixed-$V$ data obtained equivalently to those in Fig.~\ref{figure1}.
}
\label{figure4}
\end{figure}

We impose simple shear of rate ${\gdot} = \partial v_x/\partial y$ via Lees-Edwards boundaries~\cite{lees1972computer}, with flow, gradient and vorticity directions $x$, $y$ and $z$. Our time unit 
$\tauI=\sqrt{\rho a^3/k}$ and  $\tauV=\eta a/k$ in dry and wet systems respectively, giving non-dimensional shear rates  $\gdot_d=\gdot\tauI$ and $\gdot_w=\gdot\tauV$.
The stress is computed by averaging the tensor product of particle-particle vectors and forces,
and rescaled as $\Sigma_{ij}=\sigma_{ij} a/k$,
so that $\mu = \Sigmas/\Sigman$, $I=\gdot_d/\sqrt{\Sigman}$ and $J=\gdot_w/\Sigma_{yy}$. Steady state data are averaged over 30 realisations measured up to 10
strain units beyond the initial transient.
Results are independent of our timestep,
$\delta t=0.01\tauI$.
The periodic box size $L_{x,y,z}\approx30a$ in Figs.~\ref{figure1},~\ref{figure4}, with elongated $L_y$ or $L_z$ specified in Figs.~\ref{figure2},~\ref{figure3}.

To model the experimental protocols  described above, we perform simulations in two different modes.
In one we fix $V$ (and so $\phi$) and ${\gdot}$,
measuring ${\Sigmas}$ and ${\Sigman}$ in steady state.
In the other we fix ${\gdot}$ and introduce a set point normal stress $\Sigman^\mathrm{SP}$, 
 at each timestep updating 
$V(t+\delta t) = V(t) + \alpha\delta t({\Sigman}(t)-\Sigman^\mathrm{SP})/{\Sigman^\mathrm{SP}}$
and checking that this maintains $\Sigman\approx \Sigman^\mathrm{SP}$  to excellent approximation.
We set $\alpha=0.1$ having verified that results are insensitive to this choice.
We then 
measure  $\phi$ and ${\Sigmas}$ in steady state.

%==========================
%==========================
\emph{Dry system.---}
%==========================
%==========================
We start with the fixed-$\Sigman$ data.
Figures~\ref{figure1}(a)-(d) show constitutive curves  $\mu(I)$ and $\phi(I)$
obtained for several fixed $\Sigman$ by increasing ${\gdot_d}$ (and so $I=\gdot_d/\sqrt{\Sigman}$) from left to right. 
Figure~\ref{figure1}(e) shows that our control scheme produces a stable ${\Sigman}$ over time.
Viewed on the coarse scale of Figs.~\ref{figure1}(a),(c), the rheology is broadly consistent with that for hard particles,
with
master curves 
for $\mu(I)$ and $\phi(I)$ following~\cite{jop2006constitutive,boyer2011unifying}. 
%==========================
%==========================
%==========================
%==========================
However, the separation of $\mu$ and $\phi$ curves on the finer scales of Figs.~\ref{figure1}(b),(d), and the anomalous $\phi$ curves in Fig.~\ref{figure1}(c) (red lines), show a breakdown of this master scaling for larger $\Sigman$. The finite stiffness $k$ in our model allows particles to be slightly compressed,
meaning the flow state is not uniquely defined by $I$ but depends separately on $\gdot_d$ and $\Sigman$.
The $\mu(I)$ curves obtained by increasing $\gdot_d$ at fixed $\Sigman$ are then non-monotonic,  
with a window of $I$ over which $\frac{\partial\mu}{\partial I}\vert_{\Sigman}<0$
(pink boxes, Fig.~\ref{figure1}),
consistent with an earlier result in 2D~\cite{degiuli2017friction}.
This suggests that flow at an imposed global $I$ in this window 
will form gradient shear bands of differing $\gdot_d$ with layer normals along $y$.
Meanwhile the $\phi(I)$ data in Fig.~\ref{figure1}(d) reveal dilation with increasing $\gdot_d$ at fixed $\Sigman$,
with $\frac{\partial \phi}{\partial I}\vert_{\Sigman} <0$.

In this regime, where the flow depends not on $I$ alone
but on both $\gdot_d$ and $\Sigman$, one can obtain constitutive curves  at fixed $\Sigman$ by varying $\gdot_d$ (as discussed so far)
or at fixed $\gdot_d$ by varying $\Sigman$. To explore the latter, we show by the purple line in Fig.~\ref{figure1}(d) all data points with a single fixed $\gdot_d$, now across data for different $\Sigman$. Doing so reveals another non-monotonicity, highlighted by the green boxes in Fig.~\ref{figure1}, with a window of $I$ for which
${\frac{\partial \phi}{\partial I}}\vert_{\gdot_d}>0$, which means that
increasing $\Sigman$ causes dilation. This will  have important implications
for the fixed-$V$  rheology, to which we now turn.

Shown in Fig.~\ref{figure1}(f) are time-dependent signals $\Sigmas(t)$ measured at a given $\gdot_d$ for three $\phi$.
%Strains up to ${\gdot}t=1$ represent the start-up transient.
Beyond the ${\gdot}t\leq1$ start-up transient,
we observe  a statistically steady $\Sigmas$ % fluctuating about a mean value
for $\phi=0.58,~0.61$. 
In contrast, at intermediate $\phi=0.596$ the stress intermittently switches between two apparently metastable values,
each sustained for $0.5-2$ strain units.
Similar phenomenology was seen in~\cite{chialvo2012bridging},
described there as fluctuating rather than bistable behavior.
Running our simulation at 
$\phi=0.596$ and imposed ${\Sigmas}$
(using an algorithm analogous to our fixed-$\Sigman$ one)
led to flow arrest at long times as reported previously~\cite{grob2014jamming,grob2016rheological},
precluding fixed-$V$ measurement of the unstable region.
To obtain a point (or points) on the stationary constitutive curve from this $\Sigmas(t)$ signal at each $\gdot_d$ and $\phi$, we produce a histogram $P(\Sigmas)$ by 
sampling at intervals of ${\gdot}t=0.01$, and fit it to a single Gaussian for low and high $\phi$,
or a sum of two Gaussians for bistable cases at intermediate $\phi$.
The locations of the maximum (or maxima) of these fits are taken as the time-averaged stress values,
shown by solid colored lines in Figs.~\ref{figure1}(g),(h).

%==========================
%==========================
%==========================
%==========================

We additionally obtain \emph{reconstructed} fixed-$V$ constitutive curves
from the fixed-$\Sigman$ data presented in Figs.~\ref{figure1}(a)-(d).
Each plotted point therein has known $\mu$, $\phi$ and $I$, with the latter measured at known $\Sigman$ and $\gdot_d$ and with $\Sigmas=\mu\Sigman$. 
We thus replot these data $\Sigmas(\gdot_d)$ and $\Sigman(\gdot_d)$ as faded colored lines in Fig.~\ref{figure1}(g),(h).
Here $\Sigman(\gdot_d)$ are horizontal lines
along which $\phi$ decreases monotonically from left to right.
Connecting points on each of these lines that have equal $\phi$, 
one obtains the solid grey and black contours in Fig.~\ref{figure1}(g),(h).
For low ${\gdot_d}$ we observe inertial
${\Sigmas}= f(\phi){\gdot_d}^2$
rheology~\cite{chialvo2012bridging,campbell1990rapid}
for
$\phi\leq0.593$
and quasistatic
${\Sigmas}= h(\phi)$
for $\phi\geq0.605$.
For large $\gdot_d$ the data tend towards intermediate ${\Sigmas}\sim {\gdot_d}^{1/2}$ behavior~\cite{chialvo2012bridging}.
This reconstruction produces constitutive curves that are
S-shaped and show DST, 
consistent with the bi-valued constitutive curves obtained for $0.596\leq\phi\leq0.599$ in the fixed-$V$ simulation.
The particle contact number grows towards its isostatic point $z\approx4$ (for our $\mu_p$) with increasing stress at fixed $V$ (data not shown). Importantly, this suggests a different DST mechanism compared to critical load models in which the isostatic point itself decreases with increasing stress~\cite{seto2013discontinuous,wyart2014discontinuous}.
Indeed, the transition reported here links inertial or viscous branches to quasistatic ones,
as opposed to linking frictionless and frictional viscous branches as in~\cite{seto2013discontinuous,wyart2014discontinuous}.

We now show that these S-shaped $\Sigman(\gdot_d)$ curves map directly to a particular feature of $\phi(I)$ in Fig.~\ref{figure1}(d).
Writing $\phi=f(\gdot_d,\Sigman)$,
one can find the slope of fixed-$V$ curves by setting
$d\phi = 0$
and using the definition of $I$ to obtain
$\frac{d\Sigman}{d\gdot_d}\vert_{\phi} = (2\Sigman/\gdot_d)\left(\frac{\partial f}{\partial I}\vert_{\Sigman}\right)/\left(\frac{\partial f}{\partial I}\vert_{\gdot_d}\right)$~\footnote{Using
$I = \gdot/\sqrt{\Sigman}$
we obtain
$\partial \gdot\vert_{\Sigman} = \sqrt{\Sigman} \partial I\vert_{\Sigman}$
and
$\partial \Sigman\vert_{\gdot} = \frac{-2\Sigman^{3/2}}{\gdot}\partial I\vert_{\gdot}$,
which when substituted into the total derivative
$d\phi = {\frac{\partial f}{\partial \gdot}}\vert_{\Sigman} d\gdot + {\frac{\partial f}{\partial \Sigman}}\vert_{\gdot} d\Sigman = 0$
yields the expression given in the main text.}.
Thus if $\frac{\partial \phi}{\partial I}\vert_{\Sigman}<0$
(blue-to-red lines, Fig.~\ref{figure1}(d)),
then
$\frac{\partial \phi}{\partial I}\vert_{\gdot_d}>0$
(purple line)
implies downward sloping fixed-$V$ curves $\Sigman(\gdot_d)$.
Indeed, the green boxes in Fig.~\ref{figure1} defined by $\frac{\partial \phi}{\partial I}\vert_{\gdot}>0$
coincide in Fig.~\ref{figure1}(g),(h) with $\frac{\partial \Sigma_{yy}}{\partial \gdot}\vert_{\phi}<0$
(and $\frac{\partial \Sigma_{xy}}{\partial \gdot}\vert_{\phi}<0$).
The same conclusion may be obtained graphically,
by taking a horizontal line through Fig.~\ref{figure1}(d) at, say, $\phi=0.598$ (dark grey line).
One now finds three intersections with the purple line,
giving three stress states for a given $\gdot_d$ at $\phi=0.598$,
implying S-shaped fixed-$V$ constitutive curves.
Importantly, the pink and green boxes do not fully overlap,
so that non-monotonic $\mu(I)$ (pink) and $\Sigmas(\gdot)$ (green)
do not imply each other.

So far, we have reported constitutive curves that are non-monotonic and shear thinning at fixed $\Sigman$, implying a pre-disposition towards gradient banding; and, in notable contrast, 
non-monotonic and shear thickening at fixed $\phi$, implying a pre-disposition towards vorticity banding.
Significantly,
fixed-$\Sigman$ simulations with $\mu_p=0$ produce monotonic $\mu(I)$ and $\phi(I)$.
We now show that the observed non-monotonicities indeed lead to the formation of 
bands in simulation boxes large enough to accommodate heterogeneity along the relevant axis. (The boxes simulated in Fig.~\ref{figure1} are too small to do so.)

Figure~\ref{figure2} shows a simulation  at fixed $\Sigman$ in a box elongated along the gradient direction, $L_y=1200a$. The stress $\Sigman$ is such that the $\mu(I)$ curve for homogeneous shear (in a smaller box) is non-monotonic, Fig.~\ref{figure2}(c). A global shear rate $\bar{\gdot}_d$  is then imposed so that $I$ lies in its negatively sloping part  (black star).
The time series of the $y$ profile of $\gdot_d$ then clearly reveals banding,
Fig.~\ref{figure2}(a).
Binning the local $I$ values
gives the histogram in Fig.~\ref{figure2}(b),
with peaks at $I$ values indicated  by the blue and red stars in Figs.~\ref{figure2}(b),(c).
These show that the low shear band is effectively jammed ($I\approx 0$)
while the high shear band lies close to
the minimum of the homogeneous $\mu(I)$. The stress values  remain spatially and temporally uniform (not shown), with $\mu$ in the banded state slightly below that in homogeneous flow,
Fig.~\ref{figure2}(c). This may be due to non-local effects, which are known to cause deviations from  homogeneous rheology that
propagate distances $\mathcal{O}(10)a$ from yield planes~\cite{bouzid2013nonlocal}.

Figure~\ref{figure3} reports a fixed-$V$ simulation
in a box elongated along the vorticity direction, $L_z=1200a$, with $\phi=0.596$ so that $\Sigma_{xy}(\gdot_d)$  is non-monotonic, Fig.~\ref{figure3}(c).
A set-point global stress $\bar{\Sigma}_{xy}$ (black star) is imposed
near its negatively sloping part via an algorithm that dynamically adjusts $\gdot_d$. 
The time series of the $z$ profile of $\Sigma_{xy}$, Fig.~\ref{figure3}(a), (and $\Sigman$, not shown) then also shows a stress band.
This is not however stationary but steadily propagates along $z$,
consistent with the balance of normal stresses
precluding stationary banding~\cite{hermes2016unsteady}.
The $\phi$ profile is uniform throughout, within the sensitivity of our measurement.
Binning the $\Sigmas$ data gives the histogram in Fig.~\ref{figure3}(b),
with peaks  at $\Sigmas$ values (pink and yellow stars) transcribed to the $\Sigmas(\gdot_d)$ representation in Fig.~\ref{figure3}(c).
This finding is contrary to a report in 2D dry systems of a homogeneous jammed state in regions where $\Sigmas(\gdot_d)$ is expected to be downward sloping~\cite{grob2014jamming}.
Propagating vorticity bands were however seen (albeit for wet systems with critical load friction) in~\cite{chacko2018dynamic}. 

%==========================
%==========================
%==========================
%==========================

%==========================
%==========================
\emph{Wet system.---}
%==========================
%==========================
We show finally that constitutive curves for wet systems have non-monotonicities counterpart to those in dry systems.
Figure~\ref{figure4}(a) shows that $\mu(J)$ curves obtained by varying $\gdot_w$ at fixed $\Sigman$ are non-monotonic for a range (albeit limited) of $\Sigman$.
The $\phi(J)$ curve measured by instead varying $\Sigman$ at fixed $\gdot_w$ is also non-monotonic  (purple line, Fig.~\ref{figure4}(b)).
This then maps to non-monotonic reconstructed S-shaped shear thickening curves $\Sigma_{xy}(\gdot_w)$ and $\Sigma_{yy}(\gdot_w)$ at fixed $V$ in Fig.~\ref{figure4}(c),(d), as for dry systems. Simulations performed actually at fixed $V$ in this regime likewise give bi-valued constitutive curves
\footnote{The range of $\phi$ for which inaccessible stress states are observed differs between the dry and wet cases,
suggesting that the lubrication forces present in the latter system slightly change $\phim$.},
with a viscous regime $\Sigmas\propto\gdot_w$ for $\phi\leq0.58$.

For dry systems, we recall that negative slope of the $\mu$ curves at fixed $\Sigman$ (pink boxes, Fig.~\ref{figure1}) does not automatically imply negative slope of the $\Sigma_{xy}(\gdot_w)$ curves at fixed $V$ (green boxes). This lack of correspondence is even more apparent in wet systems: little overlap is apparent between the pink and green boxes in Fig.~\ref{figure4}. 
Instead, as in dry systems, the presence of negative slope in the constitutive curves $\Sigma_{xy}(\gdot_w)$ at fixed $V$  requires only that the material is dilatant with respect to increases in both $\gdot_w$ and $\Sigman$ under fixed $\Sigman$.
\citet{grob2014jamming} give a criterion for the existence of such points based on a simple model of additive stress contributions to ${\gdot}$.

%==========================
%==========================
\emph{Conclusion.---}
%==========================
%==========================
In this Letter, we have mapped the fixed-$\Sigman$ and fixed-$V$ constitutive curves of wet and dry granular flows, shown the mapping between them, and demonstrated
their connection to shear banding.
Our results highlight the shortcomings of current constitutive models:
predictions for S-shaped $\Sigmas(\gdot)$ exist, but do not predict non-monotonicity in $\mu$~\cite{nakanishi2012fluid,grob2014jamming,wyart2014discontinuous} or incorporate quasistatic flow.
A phenomenological model encoding flow-induced noise predicts non-monotonic $\mu(I)$~\cite{degiuli2017friction}, though the demonstration of non-monotonic $\mu(J)$ calls for an equivalent mechanism to be identified in overdamped systems.
Further,
our results challenge the present consensus by showing that DST can arise in the absence of a critical load model~\cite{seto2013discontinuous,wyart2014discontinuous}, with finite stiffness alone providing the requisite stress scale.
This informs an ongoing debate on whether contacts of few asperities render deformations relevant even for grains of large modulus~\cite{tabor1981friction,lobry2019shear,more2021unifying,papadopoulou2020shear}.
Reconciling the rich banding dynamics reported here with a detailed mechanistic description accounting also for non-locality~\cite{bouzid2013nonlocal,mowlavi2021interplay} and boundary effects~\cite{hu2024nonmonotonic} is an open challenge.
Relating underlying constitutive curves to measured flow curves is a longstanding problem in complex fluids, reinvigorated by our demonstration here that the measurement protocol can radically change the observed phenomenology. Further understanding the micromechanics at play is a fundamental challenge to statistical and soft matter physics, and to developing rheological constitutive models crucial to predicting macroscopic engineering flows.

\begin{acknowledgments}
The authors thank M. Cates and Y. Jiang for discussions. 
This project has received funding from the European Research
Council (ERC) under the European Union’s Horizon
2020 research and innovation programme (grant agreement No. 885146) (SMF); and from the
Royal Academy of Engineering under the Research Fellowship scheme
and from the Leverhulme Trust under Research Project Grant RPG-2022-095 (CN). 
\end{acknowledgments}

\bibliographystyle{apsrev4-1}
\bibliography{apssamp}
\end{document}